%% file: Manuscript.tex
\documentclass[%
superscriptaddress,
preprint,
 amsmath,amssymb,
 aps,
prb,
]{revtex4-2}
\usepackage{graphicx}
\usepackage{dcolumn}
\usepackage{bm}
\usepackage{subfigure}
\usepackage{color}

\newcommand{\FeCo}[2]{Fe$_{\mathrm{#1}}$Co$_{\mathrm{#2}}$}

\begin{document}
\title{Magnetization process in epitaxial Fe$_{85}$Co$_{15}$ thin films\\ via anisotropic magnetoresistance}

\author{L. Saba}
\email{lautaro.saba@ib.edu.ar}
\affiliation{Instituto Balseiro, Universidad Nacional de Cuyo (UNCuyo), Comisión Nacional de Energía Atómica (CNEA), R8402AGP San Carlos de Bariloche, Río Negro, Argentina}

\author{J. E. Gómez}
\affiliation{Instituto de Nanociencia y Nanotecnología (CNEA - CONICET), Nodo Bariloche, Av. Bustillo 9500, (8400) Bariloche, Rio Negro, Argentina}

\author{D. J. Pérez-Morelo}
\affiliation{Instituto Balseiro, Universidad Nacional de Cuyo (UNCuyo), Comisión Nacional de Energía Atómica (CNEA), R8402AGP San Carlos de Bariloche, Río Negro, Argentina}
\affiliation{Instituto de Nanociencia y Nanotecnología (CNEA - CONICET), Nodo Bariloche, Av. Bustillo 9500, (8400) Bariloche, Rio Negro, Argentina}

\author{S. Anguiano}
\affiliation{Instituto Balseiro, Universidad Nacional de Cuyo (UNCuyo), Comisión Nacional de Energía Atómica (CNEA), R8402AGP San Carlos de Bariloche, Río Negro, Argentina}
\affiliation{Instituto de Nanociencia y Nanotecnología (CNEA - CONICET), Nodo Bariloche, Av. Bustillo 9500, (8400) Bariloche, Rio Negro, Argentina}

\author{D. Velázquez Rodriguez}
\affiliation{Instituto de Nanociencia y Nanotecnología (CNEA - CONICET), Nodo Bariloche, Av. Bustillo 9500, (8400) Bariloche, Rio Negro, Argentina}

\author{A. Butera}
\affiliation{Instituto Balseiro, Universidad Nacional de Cuyo (UNCuyo), Comisión Nacional de Energía Atómica (CNEA), R8402AGP San Carlos de Bariloche, Río Negro, Argentina}
\affiliation{Instituto de Nanociencia y Nanotecnología (CNEA - CONICET), Nodo Bariloche, Av. Bustillo 9500, (8400) Bariloche, Rio Negro, Argentina}

\author{M. Granada}
\affiliation{Instituto Balseiro, Universidad Nacional de Cuyo (UNCuyo), Comisión Nacional de Energía Atómica (CNEA), R8402AGP San Carlos de Bariloche, Río Negro, Argentina}
\affiliation{Instituto de Nanociencia y Nanotecnología (CNEA - CONICET), Nodo Bariloche, Av. Bustillo 9500, (8400) Bariloche, Rio Negro, Argentina}

\author{L. Avilés-Félix}
\email{luis.aviles@ib.edu.ar}
\affiliation{Instituto Balseiro, Universidad Nacional de Cuyo (UNCuyo), Comisión Nacional de Energía Atómica (CNEA), R8402AGP San Carlos de Bariloche, Río Negro, Argentina}
\affiliation{Instituto de Nanociencia y Nanotecnología (CNEA - CONICET), Nodo Bariloche, Av. Bustillo 9500, (8400) Bariloche, Rio Negro, Argentina}

\date{\today}

\begin{abstract}
The effects of the crystalline symmetry on the magnetotransport properties in ferromagnetic alloys are being reexamined in recent years particularly due to the role of the anisotropic magnetoresistance on the electrical detection of magnetization dynamics, which is relevant to estimate spin transport parameters such as the spin Hall angle or the damping constant. In this work we investigated the crystalline dependent anisotropic magnetoresistance in epitaxial \FeCo{85}{15} films and discuss the magnetization process through the magnetotransport properties by varying the relative orientations between the electric current, the external magnetic field and the \FeCo{85}{15} crystallographic directions. We have found that the anisotropic magnetoresistance ratio depends on the current direction with respect to the crystal axes of \FeCo{85}{15} and determine a ratio of 0.20 \% and 0.17 \% when the current is applied along the [110] hard and [100] easy axes, respectively. We fit our experimental data using the Stoner-Wohlfarth model to describe the path followed by the magnetization during the magnetization process and to extract the anisotropy constants. The fitted cubic and uniaxial anisotropy constants are $K_c$ = 21 kJ/m$^3$ and $K_u$ = 11 kJ/m$^3$, which are comparable with reported values from the angular variation of ferromagnetic resonance experiments. Our results contribute to the understanding of the interplay between the crystalline structure and the magnetotransport properties of FeCo alloys.
\end{abstract}
\maketitle

\section{\label{sec:Intro}Introduction}

In the last years, \FeCo{100-x}{x} alloys with bcc crystalline structure have regained significant attention as potential candidates for the development of spintronics devices. While the presence of both Fe and Co offers an ideal combination of high magnetic saturation and tunable magnetic anisotropy, it is their ultra low damping values which are key for applications in magnetic random-access memories and spin transfer torque-based devices. Fe-rich \FeCo{100-x}{x} is the metallic ferromagnet with one of the lowest reported damping values. Since its prediction by Mankovsky \textit{et al}. \cite{Mankovsky2010} and its experimental verification by Schoen \textit{et al}. \cite{Schoen2016} , several works on the spin transport and magnetotransport properties of polycrystalline and epitaxial \FeCo{100-x}{x} thin films have been reexamined extensively \cite{Berger1988,Ganguly2014,Haidar2015,Weber2019,Zeng2020}. Some recent works have highlighted the observation of an anisotropic damping in epitaxial \FeCo{100-x}{x} films \cite{Li2019, Zeng2020a,Velazquez2024}. The origin of this anisotropic damping has been attributed to the variation of the spin-orbit coupling (SOC) along different directions of the cubic lattice \cite{Li2019}. At this point, it is also important to recall that the SOC induces a mixing of spin-up and spin-down $d$ states which depends on the magnetization direction. Therefore, the magnetization direction determines the density of unoccupied $d$ states at the Fermi level \cite{Kokado2012}. This difference in the $s$-$d$ electron scattering cross-section is the responsible for the anisotropic magnetoresistance (AMR) in ferromagnetic metals \cite{Thomson1856}. Experimentally, the AMR manifests as the dependence of electrical resistance on the relative angle between the direction of the applied current and the magnetization. Understanding AMR is crucial for quantifying damping, as it plays a key role in experimental techniques used to investigate magnetization dynamics. For instance, electrical detection of magnetization dynamics via rectification effects is employed to determine the spin torque efficiency of ferromagnetic/non-magnetic bilayers, such as in spin-torque ferromagnetic resonance experiments \cite{Harder2016,Zeng2020a,Gonzalez2022}. A clear separation of the AMR contribution can provide insights into how to disentangle the contributions of the crystalline structure from the angular dependence of in-plane spin rectification signals, thereby allowing for a more accurate quantification of the spin transport parameters.

The angular dependence of the AMR in polycrystalline magnetic materials can be deduced from the generalized Ohm's law \cite{1960Juretschke,Harder2016}. Considering a magnetic layer with in-plane magnetization and that the AMR dominates the magnetoresistance effects in ferromagnets, we can express the induced electric field in the layer as: 
\begin{equation}
\mathbf{E} = \rho_\perp \mathbf{j} + (\rho_\parallel - \rho_\perp)\mathbf{m}\left(\mathbf{m} \cdot \mathbf{j}\right), 
\label{eq:e}
\end{equation}
where $\mathbf{j}$ is the current density, $\mathbf{m}$ is the unit vector in the direction of magnetization, and $\rho_\parallel$ and $\rho_\perp$ are the resistivities when the current is parallel and perpendicular to the magnetization direction, respectively. Generally, AMR is defined as
\begin{equation}
\frac{\Delta \rho}{\rho_\perp} = \frac{\rho_\parallel - \rho_\perp}{\rho_\perp}.
\label{Eq:AMR}
\end{equation}

On the other hand, in ferromagnetic single crystals or in epitaxial films the electron scattering strongly depends on the current orientation with respect to the crystallographic directions, resulting in a distinct behavior of the AMR. Since both damping and anisotropic magnetoresistance depend on the SOC, a crystalline-dependent AMR is expected in this alloy. The experimental evidence of this crystalline dependent AMR produced by the anisotropy in the spin-orbit coupling has been reported by Y. Li \textit{et al} in \FeCo{50}{50}. \cite{Li2019}

In this paper, we study the effect of crystal symmetry on the magnetotransport properties of epitaxial \FeCo{85}{15}  films, focusing on the anisotropic magnetoresistance and its dependence with the electric current and external magnetic field directions. Epitaxial growth of FeCo films on MgO substrates provides precise control over the crystal structure, enabling us to explore the role of crystalline structure and anisotropic spin-orbit coupling in magnetotransport. We investigate the correlation between the magnetization process and the crystalline structure by varying the directions of current and magnetization with respect to different crystallographic axes. By analyzing the angular dependence of magnetoresistance and fitting the data using the generalized Ohm's law and free energy minimization, we estimated the uniaxial and cubic anisotropy constants and determine the longitudinal and transverse voltage dependence on the current and field orientations.

\section{\label{sec:Experimental} Experimental Details}

\begin{figure}[ht]
\centering
\includegraphics[width=0.5\linewidth]{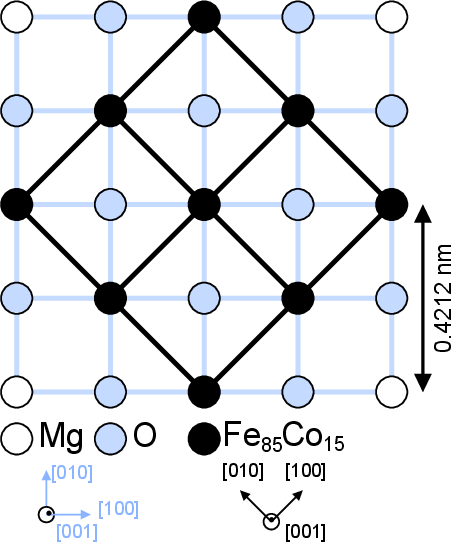}
\caption{Schematic representation of the atomic arrangement of \FeCo{85}{15} grown epitaxially on MgO. The crystallographic directions of the MgO substrate and of the \FeCo{85}{15} film are indicated at the bottom in light blue and in black, respectively. The [100] direction of the \FeCo{85}{15} film is rotated 45$^\circ$ with respect to the [100] of the MgO.}
\label{fig:cristal}
\end{figure}

An Fe$_{85}$Co$_{15}$/Pt bilayer was deposited by magnetron sputtering on MgO [001] single crystal substrates. The Fe$_{85}$Co$_{15}$ layer was deposited from an alloyed target with a substrate-to-target distance of 7.5 cm. The base pressure before depositing the samples was $\lesssim$ 10$^{-6}$ Torr and the deposition was performed under an Ar pressure of 1.8 mTorr, at a fixed power density of 1.32 W/cm$^{2}$. To improve the epitaxial growth of the FeCo the substrate was kept at a fixed temperature of 423 K during deposition. The Pt layer, on the other hand, was deposited at room temperature, under an Ar pressure of 2.6 mTorr and a power density of 1.76 W/cm$^{2}$. The nominal layer thickness was fixed to 15 nm for \FeCo{85}{15} and 10 nm for Pt. The structural and compositional characterization via X-ray diffraction and X-ray reflectivity measurements, Scanning Transmission Electron Microscopy with High Angular Annular Dark Field (STEM-HAADF) and Electron energy-loss spectroscopy (EELS) can be found in our previous works \cite{Velazquez2020,Velazquez2021,Velazquez2024}. The magnetotransport measurements were performed in a setup consisting on a Keithley 6221 current source and two HP34401A multimeters. Samples were mounted inside a Janis SVT-300 cryostat to control the temperature of the sample at 150 K. The external magnetic field was applied using a Danfysik System 8000 power supply connected to a GMW electromagnet mounted on a goniometer that can rotate 360 degrees. The sample holder was oriented such that the magnetic field can rotate within the plane of the film. The axes corresponding to the crystalline directions of the \FeCo{85}{15} layer are schematically shown in Fig. \ref{fig:cristal}. The longitudinal and transverse voltages were measured simultaneously in Fe$_{85}$Co$_{15}$/Pt micro-sized Hall bars made using standard optical lithographic techniques. The details of the Hall bars, such as the separation between the contacts and dimensions of the bar, can be found in Fig. S1 of the Supplementary Data file. Two Hall bars were used for the characterization of the magnetotransport properties, one oriented along the direction of the \FeCo{85}{15} magnetization easy axis [100] and the other bar oriented along the magnetization hard axis [110]. To determine the anisotropic magnetoresistance ratio, the longitudinal $V_{xx}$ and transverse $V_{xy}$ voltages were measured by varying the field direction ($\phi_H$), while controlling the sample temperature and the external field amplitude at 150 mT. Finally, the measurements of $V_{xx}$ and $V_{xy}$ were performed by varying the field amplitude at different angles $\phi_H$ of the external magnetic field.

\section{\label{sec:Results}Results}
Our \FeCo{85}{15} films grown on single-crystalline MgO (001) substrates exhibit cubic crystal symmetry with a bcc structure. By heating the MgO substrate during the growth of the Fe$_{85}$Co$_{15}$ film we achieve layers with the [100] axis rotated by 45$^\circ$ with respect to the MgO [100] substrate direction, as schematically shown in Fig. \ref{fig:cristal}. High-resolution transmission electron microscopy and $\phi-$scan X-ray diffraction measurements, which verify the epitaxial growth of the FeCo film, are reported elsewhere \cite{Velazquez2024}. It is important to notice that the \FeCo{85}{15} hard axes [110] and [1$\overline{1}$0] correspond to the [010] and [100] directions of the MgO substrate, respectively. 
The average value of the saturation magnetization of the \FeCo{85}{15} layers grown using the conditions mentioned above is $M_s$ = 1700 kA/m \cite{Velazquez2020,Velazquez2024}. Kerr hysteresis loops with the external magnetic field applied along different crystallographic directions of the \FeCo{85}{15} layer ([1$\overline{1}$0],[0$\overline{1}$0],[110] and [100]) confirmed that the sample presents a fourfold symmetry with the presence of a relatively weak uniaxial anisotropy and that the saturation field is around 25 mT when the external magnetic field is applied close to the magnetization hard axes. These hysteresis loops are shown in Fig S2 of the Supplementary Data file.

\subsection{\label{sec:AMR}Anisotropic magnetoresistance}
\begin{figure}[ht]
\centering
\includegraphics[width=\linewidth]{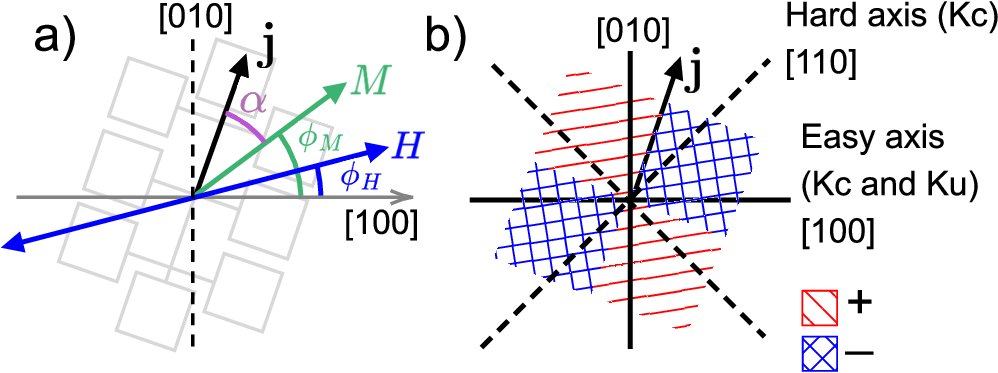}
\caption{a) Magnetization and magnetic field vectors and their respective angles for an arbitrary current direction relative to the crystalline axes of the [100] direction of \FeCo{}{}. A schematics of the Hall bar is indicated in grey. b) Representation of the quadrants in which $\rho_{xy}$ takes positive (red regions) or negative (blue regions) values, considering an arbitrary direction of the current density \textbf{j}. The crystallographic directions of the easy [100] and hard [110] magnetization axes due to cubic magnetocrystalline anisotropy are also indicated in solid and dashed lines, respectively.}
\label{fig:Angles}
\end{figure}

Two Hall bars, one bar oriented along the direction of the \FeCo{85}{15} magnetization easy axis ([100] direction in Fig. \ref{fig:cristal}) and the other bar aligned with the magnetization hard axis ([110] direction in Fig. \ref{fig:cristal}) were used for the characterization of the magnetotransport properties. From Eq. \ref{eq:e} we can express the longitudinal and transverse resistivities as
\begin{eqnarray}
\rho_{xx} = \rho_\perp + (\rho_\parallel - \rho_\perp)\cos^2\alpha, \label{Eq:rhoxx} \\
\rho_{xy} = (\rho_\parallel - \rho_\perp)\cos\alpha \sin\alpha, \label{Eq:rhoxy}
\end{eqnarray}
where $\alpha$ is the angle between the magnetization and the electric current. At this point it is important to define the angles and directions involved in the expression for $\rho_{xx}$ and $\rho_{xy}$. These parameters will help to understand the experimental results and will be used in the proposed model to determine the reorientation of the magnetization during the magnetization process. In Fig. \ref{fig:Angles}a we show a schematics of the Hall bar in which we indicate the magnetization, external magnetic field and current density vectors in the plane formed by the [100] and [010] directions of the \FeCo{85}{15} layer.

The angles $\phi_M$ and $\phi_H$ are the angles of the magnetization and magnetic field vectors measured from the \FeCo{85}{15} [100] direction, as indicated in Fig. \ref{fig:Angles}a. In Fig. \ref{fig:Angles}b we also show a representation of the quadrants in which $\rho_{xy}$ takes positive or negative values, considering an arbitrary direction of the current density vector \textbf{j} and its transverse direction. From Eq. \ref{Eq:rhoxy} we can deduce that $\rho_{xy} \propto M_{\parallel j}M_{\perp j}$, where $M_{\parallel j}$ and $M_{\perp j}$ are the components of the magnetization parallel and perpendicular to the current direction. We highlight these regions to identify the sign change of $\rho_{xy}$ during the magnetization process, which will be discussed in the forthcoming sections. The directions of the magnetization easy [100] and hard [110] axes of the \FeCo{85}{15} film resulting from the cubic magnetocrystalline anisotropy and the easy axis associated to the uniaxial anisotropy are also indicated in Fig. \ref{fig:Angles}b, where $K_c$ and $K_u$ are the anisotropy constants associated to the cubic and uniaxial anisotropies, respectively. 

\begin{figure}[ht]
\centering
\includegraphics[width=\linewidth]{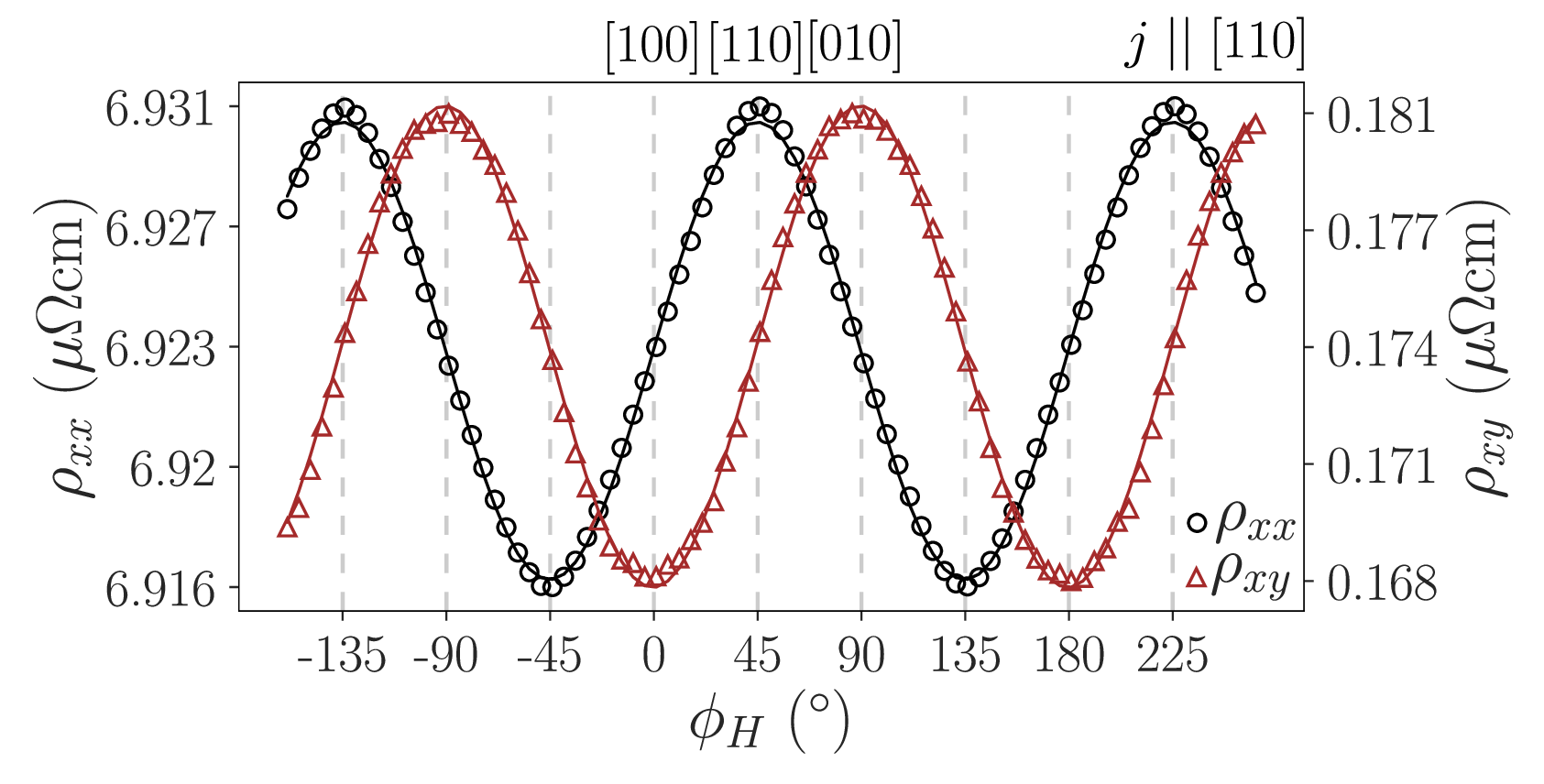}
\caption{Longitudinal ($\rho_{xx}$) and transverse ($\rho_{xy}$) resistivities as a function of the applied field angle measured at T = 150 K with an external magnetic field of 150 mT and a current density \textbf{j} = $10^6$ mA/cm$^2$. Red triangles and line correspond to the experimental data and fit of $\rho_{xy}$, respectively and the black circles and line correspond to the experimental data and fit of $\rho_{xx}$. The current is applied along the FeCo [110] (hard) direction.}
\label{fig:AMR010}
\end{figure}

To determine the AMR ratio, measurements of the longitudinal ($V_{xx}$) and transverse ($V_{xy}$) voltages relative to the current direction were performed by varying the field direction in the plane of the sample. Fig. \ref{fig:AMR010} shows $\rho_{xx}$ and $\rho_{xy}$ as a function of $\phi_H$ measured at 150 K. A current density \textbf{j} = 10$^6$ mA/cm$^2$ was applied along the [110] direction, and the applied field was 150 mT. This value of magnetic field is high enough to saturate the \FeCo{85}{15} magnetization ($\phi_M = \phi_H$) for all in-plane orientations. 

As expected, $\rho_{xx}^{[110]}$ is maximum due to \textbf{j}$\parallel$[110], and decreases as we rotate the external magnetic field towards the directions transverse to the current. On the other hand, the transverse resistivity curve ($\rho_{xy}$) is shifted by 45$^\circ$ as expected from the $\sin\alpha\cos\alpha$ term of Eq. \ref{Eq:rhoxy}. To estimate the AMR ratio the data were fitted using the expressions for $\rho_{xx}$ and $\rho_{xy}$ shown in Eqs. \ref{Eq:rhoxx} and \ref{Eq:rhoxy}. To obtain the AMR ratio from the $\rho_{xy}$ curve we corrected the respective offset in Fig. \ref{fig:AMR010}. From the fitting we obtained an anisotropic magnetoresistance ratio AMR$^{\textbf{j} \parallel [110]}$ = 0.20 $\pm$ 0.01 \%. The angular dependence of the resistivity with the applied current along the FeCo [100] direction (easy axis) was also measured. In this case, we obtained AMR$^{\textbf{j}\parallel\mathrm{[100]}}$ = 0.17 $\pm$ 0.01 \%. These values are of the same order as those found in the literature. R. P. van Gorkom \textit{et al}. \cite{vanGorkom2001} reported a value close to 0.5 \% for epitaxial Fe films at 150 K, while a more recent work by Zeng \textit{et al}. \cite{Zeng2020a} reported a strong AMR variation in epitaxial \FeCo{50}{50} films (20 nm) of 0.2 \% and 2.1 \% when the current is applied along the [110] and [010] directions, respectively, which is also correlated to a strong anisotropic damping (up to 520 \%). In our case, we have reported that although the damping along easy and hard magnetization axes is the same, $\alpha_\mathrm{easy} = \alpha_\mathrm{hard}$ = 2.6 $\times$ 10$^{-3}$ \cite{Velazquez2024}, but we can not rule out that the epitaxial \FeCo{85}{15} presents some degree of anisotropic damping associated to the AMR ratio variations. Following this reasoning we would expect a larger variation of the AMR in \FeCo{}{} alloys with more Co content. Our measurements of AMR ratio along [100] and [110] confirm that the AMR depends on the crystallographic direction in which the current is injected. The crystalline-dependent AMR can be explained by the different electronic structures along the different crystallographic directions. This phenomenon has been studied over the last years and some of the AMR ratios measured in epitaxial \FeCo{}{} alloys presented variations of $\approx$ 100 \%. Some of these values are listed in Table \ref{tab:AMR}. 

\begin{table}[ht]
    \caption{Reported AMR ratios in \FeCo{100-x}{x} alloys}
    \centering
    \begin{tabular}{|cccc|}
        \hline
        Alloy & AMR & Fabrication & Reference \\
        \hline
        \hline
        Fe$^a$              &  0.14 \% & MBE & Tondra \textit{et al}. \cite{Tondra1993}\\
        Fe$^b$              &  0.07 \% & MBE & Tondra \textit{et al}. \cite{Tondra1993}\\
        Fe$^c$              &  0.22 \% & MBE & Tondra \textit{et al}. \cite{Tondra1993}\\
        \FeCo{100-x}{x}, x=0-65$^d$ & 2.4 \% & MBE& Zeng \textit{et al}. \cite{Zeng2020}\\
        \FeCo{85}{15}$^e$   & 0.20 \% & Sputtering & This work \\
        \FeCo{85}{15}$^f$   & 0.17 \% & Sputtering & This work \\
        \FeCo{50}{50}       & 0.18 \% & Sputtering & Haidar \textit{et al}. \cite{Haidar2015}\\
        \FeCo{50}{50}$^g$   & 0.2 \%  & MBE & Zeng \textit{et al}. \cite{Zeng2020a}\\
        \FeCo{50}{50}$^h$   & 2.1 \%  & MBE &  Zeng \textit{et al}. \cite{Zeng2020a}\\
        \FeCo{25}{75}       & 0.07 \% & Sputtering & Haidar \textit{et al}. \cite{Haidar2015}\\
        \hline
    \end{tabular}\\
       \footnotesize $^a$\textbf{j}$\parallel$[001] $^b$\textbf{j}$\parallel$[110] $^c$\textbf{j}$\parallel$[111] $^d$Maximum value obtained for \FeCo{62}{38} $^e$150 K, \textbf{j} parallel to hard axis $^f$150 K, \textbf{j} parallel to easy axis $^g$\textbf{j} parallel to \FeCo{50}{50} easy axis $^h$\textbf{j} parallel to \FeCo{50}{50} hard axis .
    \label{tab:AMR}
\end{table}

\subsection{Magnetization process from the magnetotransport measurements}
\begin{figure}[ht] 
\centering
\includegraphics[width=0.5\linewidth]{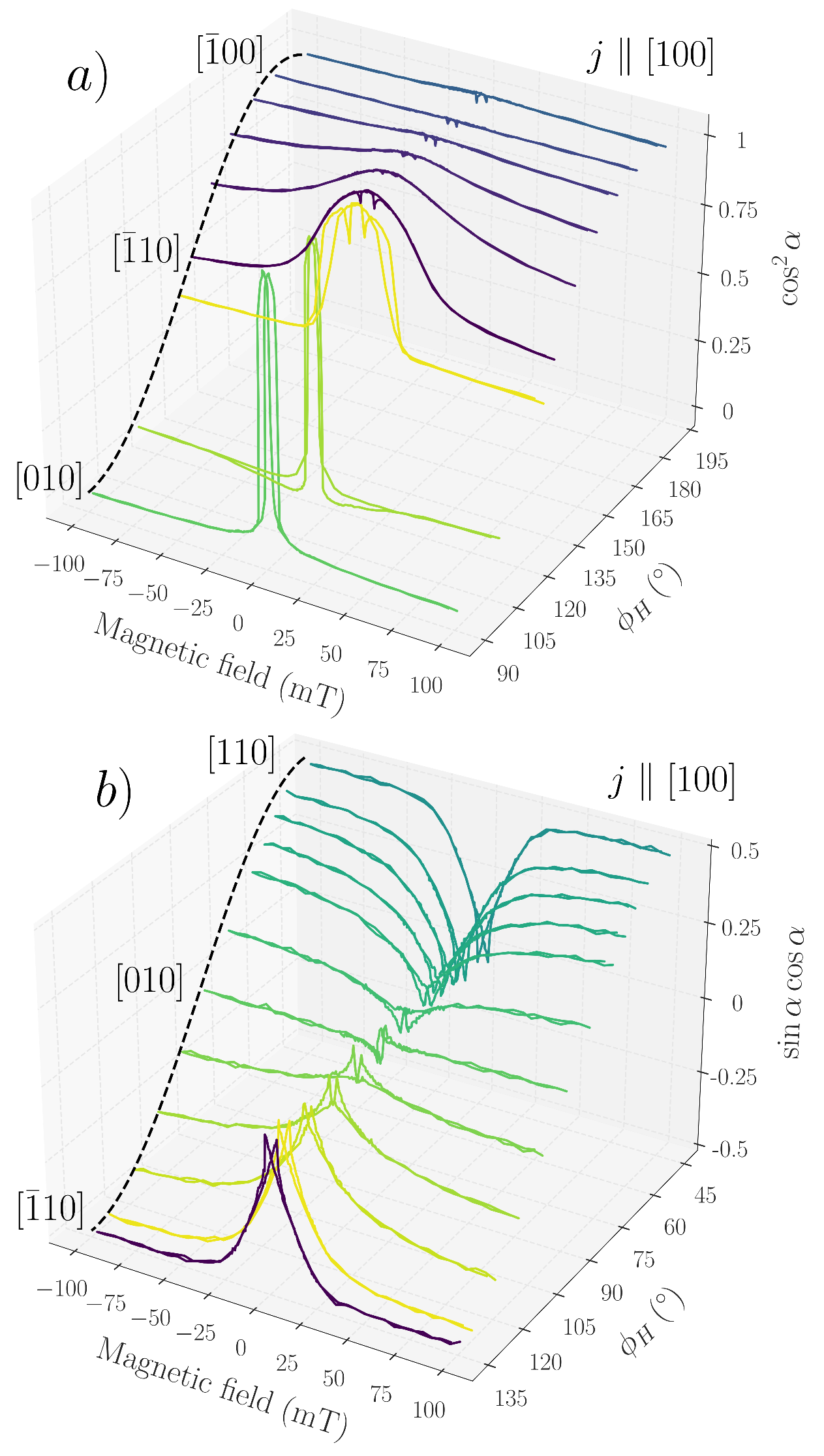}
\caption{Normalized longitudinal and transverse voltages as a) $\cos^2\alpha$ and b) $\sin\alpha\cos\alpha$ as a function of the magnetic field along different directions ($\phi_H$). $\alpha$ is the angle between the magnetization and the current, which was applied along the [100] direction. The dotted lines show the $\phi_H$ dependence of $\rho_{xx}$ and $\rho_{xy}$ when the magnetization is saturated. Measurements performed with the external magnetic field oriented along the easy or hard magnetization axes are identified with the crystallographic directions [$\overline{1}$00] and [110], respectively.}
\label{fig:4}
\end{figure}

From the longitudinal and transverse voltages, we studied the magnetization process when applying the external magnetic field along different in-plane directions ($\phi_H$). Figs. \ref{fig:4}a and \ref{fig:4}b show the normalized longitudinal and transverse voltages as a function of the external magnetic field along different directions, with angles $\phi_H$ measured from the [100] direction (Fig. \ref{fig:Angles}).
The dotted line in Fig. \ref{fig:4}a allows us to identify the evolution of the longitudinal resistance when the external magnetic field is applied from [$\overline{1}$00] to [010] (Fig. \ref{fig:AMR010}). When the external field is applied along [$\overline{1}$00] ($\phi_H$ = 180$^\circ$) or along its equivalent direction [100] ($\phi_H$ = 0$^\circ$), $\rho_{xx}$ is almost constant and independent of the magnetic field value since [$\overline{1}$00] corresponds to the \FeCo{85}{15} easy axis and the magnetization remains along [$\overline{1}$00] even at $H$ = 0. In contrast, when the external magnetic field is applied along [$\overline{1}$10] ($\phi_H$ = 135$^\circ$), the longitudinal resistance $\rho_{xx}$ increases as we decrease the magnetic field. This is because [$\overline{1}$10] corresponds to a hard axis and thus the magnetization moves away from that direction towards the easy axis as the magnetic field decreases, giving a greater resistance. 
For the range $90^\circ < \phi_H < 135^\circ$, a sharp maximum appears in the normalized $\rho_{xx}$ curve close to the coercive field value and the transition between low and high resistance states becomes abrupt as we approach $\phi_H$ = 90$^\circ$. The high resistance state observed at $H$ = 0 corresponds to the reorientation of the magnetization from [010] to [100] at low magnetic field values originated by the presence of the uniaxial anisotropy parallel to [100]. In Fig. \ref{fig:4}b the normalized transverse voltage is also plotted as a function of the external magnetic field along different directions. In this figure we plotted the curves from $\phi_H$ = 45$^\circ$ to $\phi_H$ = 135$^\circ$. Unlike the longitudinal voltage $V_{xx}$, which is proportional to the square of the component of the magnetization along the current direction, the transverse voltage contains information of the product of the components parallel and perpendicular to the current direction. Thus, when the external magnetic field is applied along [010] ($\phi_H$ = 90$^\circ$) the transverse signal remains nearly constant for both positive and negative values of the external magnetic field. This is because $\phi_H$ = 90$^\circ$ corresponds to an easy axis associated to the cubic anisotropy and the small signal observed close to $H$ = 0 is the immediate reorientation of the magnetization. As expected, the transverse voltages measured at $\phi_H$ = 45$^\circ$ and $\phi_H$ = 135$^\circ$ are opposite in sign since in both cases we start with the magnetization oriented along a different hard axis. When the external field is applied along [$\overline{1}$10], the increase of the transverse voltage for low magnetic field values is associated to the rotation towards the [010] direction and can be explained by a decrease of the components along the current direction and an increase of the component transverse to the current direction. 
The magnetization process and the model proposed to track the direction of the magnetization with respect to the crystalline axes and to the Hall bar orientation will be discussed in detail in the following subsection.

\subsection{Model}
\label{sec:Model}
The magnetization process of the \FeCo{85}{15} film was studied considering a macroscopic single domain approximation as in the Stoner-Wohlfarth model. Within this approach, we can consider that the magnetization rotates coherently for high magnetic field values although the existence of magnetic domains below the coercive field could not be neglected. To propose an expression of the free energy density we consider the shape anisotropy, an in-plane uniaxial easy axis and a cubic magnetocrystalline energy. We express this energy density using spherical coordinates ($\theta$, $\phi$) for the representation of the magnetization \textbf{M} and external magnetic field \textbf{H} vectors as shown in Fig. \ref{fig:Angles}. The magnetic free energy of the system is then described as follows:
\begin{equation}
\begin{aligned}
\label{Eq.Energy}
F(\theta_M, \phi_M) = -MH \sin \theta_M \cos(\phi_M - \phi_H)\\ + 2\pi M^2 \cos^2 \theta_M - \frac{K_c}{4} (\sin^2 2\theta_M + \sin^2 2\phi_M \sin^4 \theta_M) \\- K_u \sin^2\theta_M \cos^2(\phi_M - \phi_u),
\end{aligned}
\end{equation}
where the first term corresponds to the Zeeman energy, the second term corresponds to the shape anisotropy, the third term to the cubic anisotropy and the last term to the uniaxial anisotropy. $K_c$ and $K_u$ are the anisotropy constants associated to the cubic and uniaxial anisotropies and are defined as $K_c = \frac{1}{2} \mu_0M_sH_c$ and  $K_u = \frac{1}{2} \mu_0M_sH_u$. In the expressions for $K_c$ and $K_u$, $H_c$ and $H_u$ are defined as the cubic and uniaxial anisotropy fields. In the model, it is also assumed that the uniaxial axis is in the plane of the sample in a direction given by the angle $\phi_u$. By differentiating with respect to $\theta_M$ and making $\frac{\partial F}{\partial \theta_M} = 0$, we obtain the equilibrium angle $\theta_M^\mathrm{eq} = \frac{\pi}{2}$. This results in a simplified free energy expression containing only terms in $\phi_M$:
\begin{equation}
\begin{aligned}
\label{eq:F}
F(\phi_M) = -MH \cos(\phi_M - \phi_H) - \frac{K_c}{4} \sin^2 2\phi_M \\- K_u \cos^2(\phi_M - \phi_u).
\end{aligned}
\end{equation}

\begin{figure}[ht]
\centering
\includegraphics[width=0.75\linewidth]{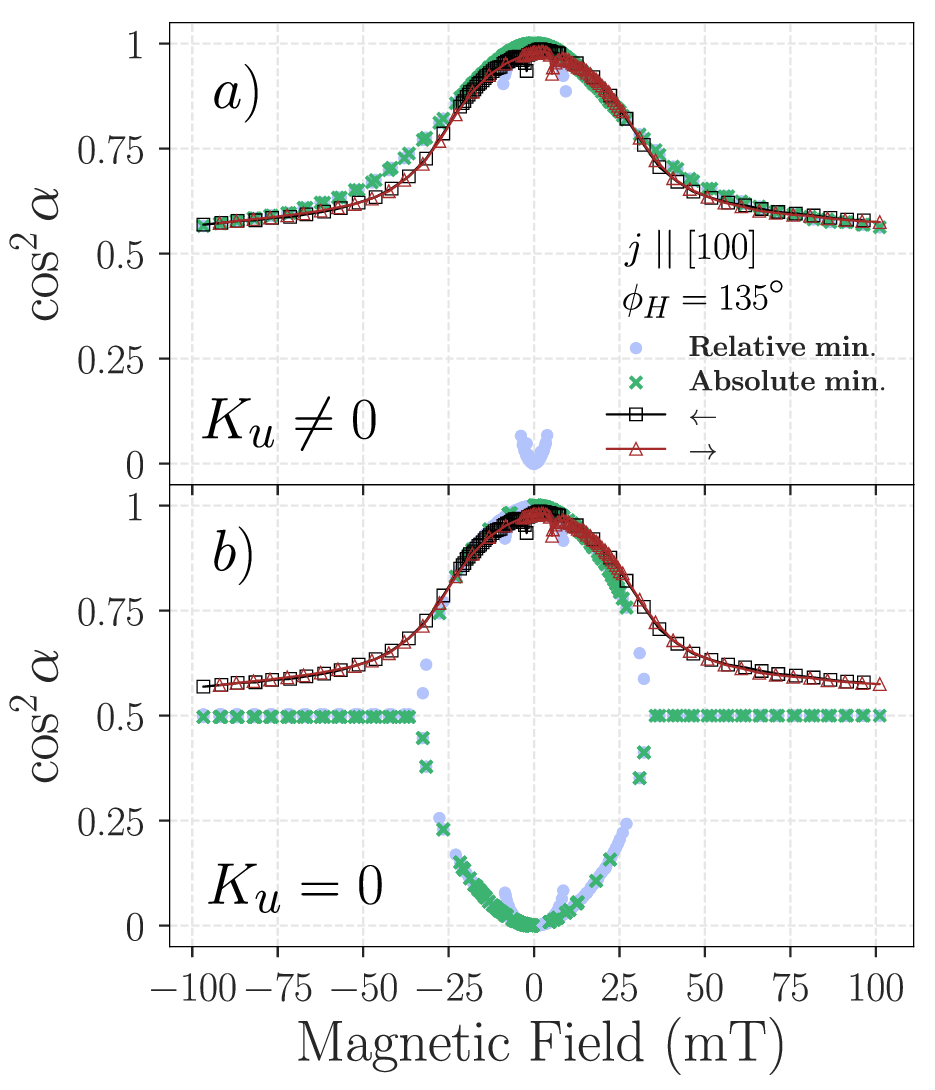}
\caption{Normalized longitudinal voltage to plot $\cos^2\alpha$ as a function of the magnetic field for $\phi_H$ = 135$^\circ$. The current was applied along [100]. The triangles and squares symbols correspond to the data measured while increasing and decreasing the magnetic field, respectively. The light blue circles correspond to predictions of the magnetization orientations that are relative energy minima. Green crosses indicate the absolute minima. Fitting was performed for a) $K_u \neq$ 0 and b) $K_u$ = 0. The fitted anisotropy constants are $K_c$ = 21 kJ/m$^3$ and $K_u = 11$ kJ/m$^3$.}
\label{fig:45}
\end{figure}

By minimizing the free energy proposed in Eq. \ref{eq:F}, we found the equilibrium angles of the magnetization ($\phi_M$) for each value of the external magnetic field $H$ and for the different directions $\phi_H$. We have considered the anisotropy constants $K_c$ and $K_u$ as fitting parameters of the normalized longitudinal and transverse voltages shown in Fig. \ref{fig:4}a and \ref{fig:4}b. Although the Kerr loops and our previous work on \FeCo{100-x}{x} alloys \cite{Velazquez2024} confirmed that the uniaxial anisotropy is weak compared to the cubic anisotropy, we need to consider it in the energy expression for the determination of the magnetization equilibrium angles, since the presence of the uniaxial anisotropy plays an important role to fit the experimental curves. This is clearly seen in Fig. \ref{fig:45} where we show the fitting of the longitudinal voltage measured with the external magnetic field at $\phi_H$ = 135$^\circ$ and the current applied along the magnetization easy axis [100]. We compare the fitting of the experimental data with $K_u \neq$ 0 (Fig. \ref{fig:45}a) and $K_u$ = 0 at $\phi_u$ = 0 (Fig. \ref{fig:45}b). The uniaxial anisotropy breaks the symmetry of the magnetization process during the application of the external magnetic field and is critical to determine the magnetization orientation, especially at low values of $\phi_H$. 

As can be seen in Fig. \ref{fig:45} the absolute minimum of the calculated $\cos^2\alpha$ can take values below 0.5. Although all angles of magnetization corresponding to a relative minimum of the free energy are plotted (light blue points in Fig. \ref{fig:45}), the points corresponding to absolute minima are distinguished. The consistency between the model and the experiment is maintained with a remarkable level of similarity in all measurements. The obtained anisotropy constants are $K_c = 21(5)$ kJ/m$^3$ and $K_u = 11(5)$ kJ/m$^3$. These values are comparable with the ones obtained from the angular variation of the ferromagnetic resonance experiments in our previous work \cite{Velazquez2024}: $H_c \sim$ 366 Oe or 29.1 kA/m, which correspond to a cubic anisotropy constant $K_c =$ 31.1 kJ/m$^3$, considering a saturation magnetization $M_s$ = 1700 kA/m.

\begin{figure}[ht]
\centering
\includegraphics[width=0.55\linewidth]{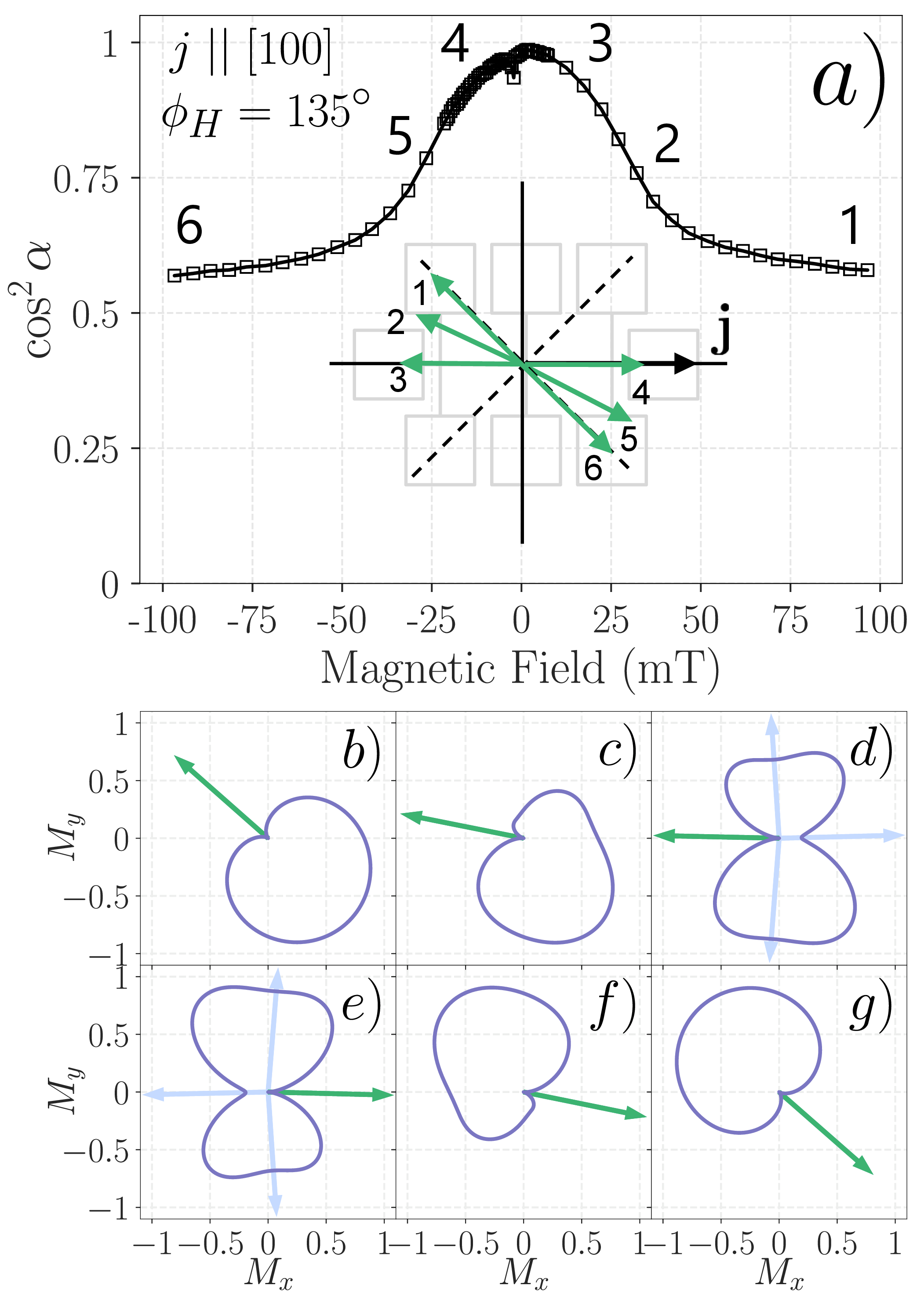}
\caption{a) Normalized longitudinal voltage to plot $\cos^2\alpha$ as a function of the magnetic field. The current was applied along the \FeCo{85}{15} [100] direction with $\phi_H$ = 135$^\circ$. Only the experimental points obtained while decreasing the field from 100 mT to -100 mT are shown. The directions of the magnetization at each stage of the measurement were identified and represented as green arrows. Inset: Schematics of the Hall bar with the current applied along the \FeCo{85}{15} [100]. b) - g) Normalized $M_x$ and $M_y$ components of the magnetization for different applied magnetic fields. b) 0.1 T, c) 0.01 T, d) 0 T, e) -0.001 T, f) -0.01 T and g) -0.1 T. The light blue arrows correspond to the direction of relative minima obtained from the minimization of the free energy. The simulated data was obtained by assuming that the $K_u$ direction is along $\phi_u$ = 0$^\circ$.}
\label{fig:6}
\end{figure}

Additionally, longitudinal and transverse voltage measurements allow us to predict the directions taken by the magnetization when the external field is applied. Fig. \ref{fig:6}a shows the normalized longitudinal voltage when the current is applied along the easy magnetization axis. In the inset we schematically represented the magnetization direction when we decrease the external magnetic field from 100 mT to -100 mT. 

In Fig \ref{fig:6}a when the magnetization is close to the saturated state at $H$ = 100 mT (green arrow 1 in the inset of Fig \ref{fig:6}a), the normalized longitudinal voltage starts around $\cos^2\alpha$ = 0.5 suggesting that the magnetization is not fully aligned with the magnetic field direction. When we decrease the external field to H = 0, the magnetization aligns with the cubic easy axis, which is parallel to the current direction (arrow 3), then the longitudinal voltage increases to $\cos^2\alpha$ = 1. As we start to increase the magnetic field in the opposite direction the magnetization abruptly rotates towards the opposite direction along the easy axis (arrows 3 and 4 in the inset). This is observed in the normalized longitudinal voltage as a small peak close to the coercive field value (around 2 mT). As we approach to saturation for negative values of the external magnetic field the magnetization smoothly rotates following the trajectories indicated by the green arrows 5 and 6 in the inset. It is important to notice that $\cos^2\alpha$ never vanishes. This indicates that the magnetization never aligns with the easy axis [010] which is perpendicular to the current direction. This is due to the presence of the uniaxial anisotropy that favors the orientations of $M$ along the [100] direction.

The description of the path followed by the magnetization during the magnetization process was verified through the determination of the magnetization equilibrium angles $\phi_H$ obtained by minimizing the free energy expression proposed in Eq. \ref{eq:F}. From Eqs. \ref{Eq:rhoxx} and \ref{Eq:rhoxy} we were able to simulate all the longitudinal and transverse signals as a function of the external magnetic field. In Figs. \ref{fig:6}b - \ref{fig:6}f we schematically show the normalized $M_x$ and $M_y$ components obtained from the minimization of Eq. \ref{eq:F} for $\phi_H$ = 135$^\circ$. To find relative minima, the energy was evaluated at 1000 equally spaced values of $\phi_M$, comparing each result with its consecutive values. This procedure was performed while keeping the anisotropy constants, as well as the magnitude and direction of the applied field, fixed. Additionally, by setting the current direction, each value of $\phi_M$ yields a corresponding $\alpha$, which is then used to compute $\cos^2\alpha$ and $\sin\alpha\cos\alpha$. The squared distances between the experimental points and the model can be easily computed in the graphs of $\cos^2\alpha$ and $\sin{\alpha} \cos{\alpha}$. To fit the anisotropy constants, we employed the Powell method to adjust these constants and minimize the residuals. In the plot we show the magnetization direction (green arrow) for the following values of the external magnetic field: b) 0.1 T, c) 0.01 T, d) 0 T, e) -0.001 T, f) -0.01 T and g) -0.1 T for $K_c/K_u \simeq 2$.

\begin{figure}[ht]
\centering
\includegraphics[width=0.55\linewidth]{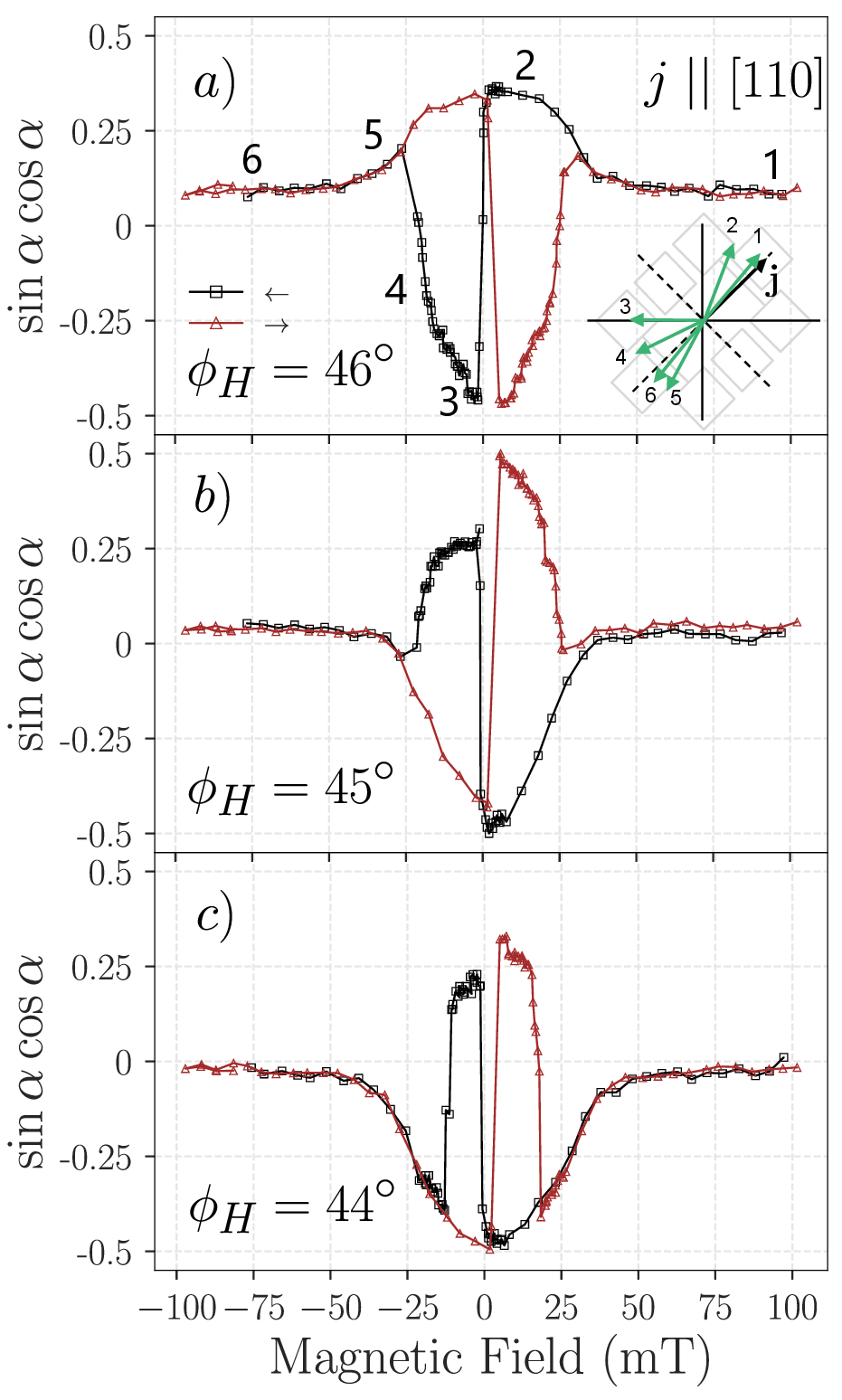}
\caption{Normalized transversal voltage to plot $\sin\alpha\cos\alpha$ as a function of the magnetic field for a) $\phi_H$ = 46$^\circ$, b) $\phi_H$ = 45$^\circ$ and c) $\phi_H$ = 44$^\circ$. The current was applied along [110]. The brown and black symbols correspond to the data obtained during increasing and decreasing the magnetic field, respectively. The light circles correspond to predictions of the magnetization orientations that are relative energy minima. Green crosses indicate the absolute minima.} 
\label{fig:phi_0}
\end{figure}

In a similar way we can correlate the magnetization process with the transverse voltage when the current is applied along the hard axis. Fig. \ref{fig:phi_0}a shows the normalized transverse voltage as a function of the external magnetic field applied along the hard axis. However, in this case the current is also applied close to the magnetization hard axis.

In the saturation state the magnetization is aligned with both the current and the magnetic field (arrow 1 in the inset of Fig. \ref{fig:phi_0}a), then the transverse voltage is close to zero. This is attributed to a misalignment between the magnetization and the external magnetic field before saturation, as can be seen in the Kerr hysteresis loops in Fig. S1. As we decrease the magnetic field to zero, the magnetization rotates to the easy axis ([010] direction of the \FeCo{85}{15}) and the normalized transverse voltage approaches to $\sin\alpha\cos\alpha$ = 0.5 (region 2 of the plot). This is because the transverse component of the magnetization with respect to the current direction increases and the longitudinal component decreases as we decrease the external magnetic field. For negative values of the magnetic field the transverse signal switches to take negative values at the coercive field. Then as we increase the external field the transverse voltage increases from $\sim$ -0.5 to 0.25 until $\sim$ 25 mT and then decreases again until the magnetization reaches the saturation for negative values of the magnetic field. The increase followed by a subsequent decrease of the transverse voltage close to 25 mT represents a rotation of the magnetization from the cubic easy axis indicated by arrow 3 in the inset of Fig. \ref{fig:phi_0}a towards an intermediate direction between the closer cubic easy axis and the direction of the external magnetic field. From this point the magnetization will coherently rotate to align with the external field in saturation. Although this magnetization process is consistent with the principle that less energy is required when the magnetization is oriented along or near the cubic easy axis in a four-fold magnetocrystalline anisotropy system, it is important to notice that the coherent rotation of the magnetization assumed in the Stoner-Wohlfarth is no longer valid. From the magnetization process described in the inset of Fig. \ref{fig:phi_0}a we would expect that, when the field is brought to zero, the magnetization would rotate towards the [100] direction (easy-easy) instead of the [010] (easy-hard) direction, however the measured transverse voltage does not reflect that. This is due to the appearance of domain walls in the \FeCo{85}{15} film as the external magnetic field approaches zero and the displacement of these domain walls  when we decrease the external magnetic field at a given $\phi_H$ \cite{Zhang2016,Buchmeier2009}. Zhang \textit{et al}. \cite{Zhang2016} demonstrated through the detection of the longitudinal and transverse signal of Kerr magnetometry in an Fe/MgO(001) single crystal, that the domain wall displacement strongly depends on the direction of the magnetic field.

Zhang \textit{et al}. also highlighted that the magnetization reversal process is very sensitive to small deviations of the external magnetic field with respect to the hard axis of Fe(001). Kerr loops shown in Ref. \cite{Zhang2016} measured at deviations of $\Delta\phi_H$ = -1$^\circ$ and $\Delta\phi_H$ = 1$^\circ$ from the cubic hard axis are determinant for the magnetization process, causing the magnetization to rotate towards [100] or [010], respectively. From our experiments and simulations we were also able to verify the major role of the relative orientation between the external magnetic field and the electric current and how this impacts on the rotation of the magnetization, similar to the analysis reported in Zhang \textit{et al}. \cite{Zhang2016} We demonstrated that a small field deviation with respect to the current direction and the cubic easy and hard axes strongly influences on the longitudinal and transverse voltages. The path taken by the magnetization is strictly related to the quadrant in which $\mathbf{H}$ was applied when it was reversed. Fig. \ref{fig:phi_0} presents the transverse signal as a function of the external field applied along a) $\phi_H = 46^\circ$, b) $\phi_H = 45^\circ$ and c) $\phi_H = 44^\circ$. These plots evidence that the magnetization process is mediated by the direction of the external field. This can be seen in the experimental data measured when we sweep the magnetic field from +100 mT to -100 mT. On the one hand, in Fig. \ref{fig:phi_0}c we can observe that the transverse voltage, which depends on the product of the $M_x$ and $M_y$ components of the magnetization, obtained for $\phi_H = 44^\circ$ decreases as we decrease the external field. However, in Fig. \ref{fig:phi_0}a the transverse voltage measured when $\phi_H = 46^\circ$ increases as we approach to zero field. The magnetization process follows a path similar to the one described in Fig. \ref{fig:6}b however the forward and return paths take different routes. In both paths, two discontinuous jumps occur, corresponding to abrupt changes in the magnetization direction that aligns near different cubic easy axes.

It is important to notice that the experimental data for magnetic field values between -50 mT $< H < $ 50 mT shown in Fig. \ref{fig:phi_0} deviate from the model used for fitting. This is commonly observed in systems in which the formation of the magnetic domains is not considered. Despite this, the magnetization process and, more important, the paths taken by the magnetization fit adequately for all the presented measurements.  

\section{Conclusions}
In summary, we have performed a detailed magnetotransport characterization in epitaxial \FeCo{85}{15} films to discuss the influence of the crystal symmetry on the anisotropic magnetoresistance. The crystalline structure produces a crystalline-dependent AMR ratio of 0.20 \% and 0.17 \% when the current is applied in the directions of a hard and easy magnetization axis respectively. These values are consistent and of the same order of magnitude than those reported in the literature and confirm that the AMR exhibits anisotropy with respect to the crystallographic direction in which the current is injected. For the FeCo studied concentration the anisotropy ratio is relatively small, which is in accordance with the reported almost isotropic values of the damping constant from FMR studies. On the other hand, the transverse and longitudinal voltages combined with a model based on the Stoner-Wohlfarth formalism allowed us to determine the magnetization direction as a function of the external magnetic field and also the identification of the paths taken when the magnetic field was reduced and then reversed. The cubic and uniaxial anisotropy constants that better fit our experiments are $K_c$ = 21 kJ/m$^3$ and $K_u$ = 11 kJ/m$^3$. Our results suggest that the \FeCo{}{}-based system opens the possibility to develop magnetoresistive sensors with tuned AMR only by designing the magnetic device oriented along a specific crystal axis.

\begin{acknowledgments}
Technical support from Rubén E. Benavides, César Pérez and Matías Guillén is greatly acknowledged. This work was partially supported by Conicet under Grant PIBAA 2022-2023 (project MAGNETS) Grant ID. 28720210100099CO; ANPCyT Grant PICT-2018-01138, ANPCyT Grant PICT 2021-00113 (project DISCO) and U.N. Cuyo Grant 06/C556 from Argentina. We acknowledge the financial support of European Commission by the H2020-MSCA RISE project ULTIMATE-I (Grant No. 101007825). L. Avilés-Félix thanks E. De Biasi for fruitful discussions and Claudio Ferrari from the Departamento de Micro y Nanotecnología (CNEA - INN) for the preparation of the lithography masks.
\end{acknowledgments}

\bibliographystyle{apsrev4-2}
\input{Manuscript.bbl}
\end{document}

%% file: Manuscript.bbl
\providecommand{\noopsort}[1]{}\providecommand{\singleletter}[1]{#1}%
%